# Upper-body musculoskeletal pain and eye strain among language professionals: a descriptive, cross-sectional study

Emma Goldsmith[*]
Madrid, Spain

## Abstract

Language professionals spend long hours at the computer, which may have an impact on their short- and long-term physical health. In 2023, I ran a survey to investigate workstation ergonomics, eye and upper-body problems, and self-reported strategies that alleviate those problems among language professionals who work sitting or standing at a desk. Of the 791 respondents, about one third reported eye problems and over two-thirds reported upper-body aches or pains in the past 12 months, with significantly higher upper-body pain prevalence among females than males, and also among younger respondents than older ones. While the pain prevalence rate in the survey was similar to figures published in the literature, as was the sex risk factor, the association of higher pain prevalence among younger people contrasted with other studies that have found increasing age to be a risk factor for pain. In this article I share the survey results in detail and discuss possible explanations for the findings.

## 1. Introduction

Sedentary behaviour is associated with poor health outcomes,[1] which is bad news for language professionals, since their work is mostly sedentary by nature. In addition, prolonged computer use can cause eye strain and pain or discomfort in the upper body[2, 3] and therefore language professionals – specifically those who facilitate written communication such as editors and translators – are an at-risk group. Because many language professionals are freelancers working from home, they need to find their own tailor-made combination of exercise, breaks and ergonomic strategies to stay healthy and prevent acute and chronic problems.

In February 2023 I gave a talk for Mediterranean Editors and Translators Association entitled *Are you sitting comfortably? Then we'll begin*. We discussed workstation set-ups and features of conventional and ergonomic computer peripherals, desks and chairs that might improve comfort. A survey completed in advance by all 40 participants helped me tailor the talk to specific upper-body musculoskeletal (MSK) problems mentioned and identify strategies that helped counteract those problems. Pain prevalence was high (81.0%), perhaps unsurprisingly since people with aches and pains might very likely be attracted to a talk on this subject.

I decided to launch the same survey on a larger scale to investigate the physical impact on language professionals of working long hours at the computer, identify self-reported strategies that alleviate that impact, and gain an overview of respondents' exercise and break habits.

---

[*] Corresponding author:
research@goldsmithtranslations.com
www.goldsmithtranslations.com

## 2. Method

I shared the survey via the EUSurvey platform on social media and through language professionals' associations worldwide over six weeks between February and April 2023. Eligible participants were full- and part-time language professionals who spend hours sitting or standing at a desk. No personally identifiable information was collected and all items were optional.

The survey had 35 items grouped into five sections:

1. General. This included demographic data on biological sex, age by decade, number of years worked and working hours per week.
2. Breaks, exercise and health. This section included the purpose and type of short breaks taken, total moderate or vigorous aerobic exercise in hours per week, muscle-strengthening activities, past or present physical discomfort in the upper body based on the Standardized Nordic questionnaire for MSK symptoms,[4] specific body areas affected and strategies to help prevent or solve problems.
3. Workstation ergonomics. This included chair, keyboard, mouse and desk types, and standing, walking or cycling workstations.
4. Monitors and eyes. After an initial question about monitor types, the rest of the section focused on eye protection strategies and past or present eye problems, defined as eye strain (tired, itchy or watery eyes), blurred vision, dry eyes and frequent headaches.
5. Strategies and accessories. These items included typing technique, mouse use and positioning, and use of other ergonomic accessories.

Two free-text fields were provided for final comments.

For the full survey (all questions and possible responses), see Supplementary File 1.

*2.1 Data analysis*

The Kruskal-Wallis test was used to compare independent variables (sex, age, years worked, hours worked, breaks and exercise) with upper-body pain and eye strain. The chi-square test was performed to identify associations between variables. I did the data analysis with JASP 18.1 and a statistician ran the data through SPSS 27 to corroborate my findings.

## 3. Results

*3.1 Demographics*

I received a total of 791 valid responses from language professionals, 81.3% of whom were female. About half (54.2%) were in their 40s and 50s. People in their 20s accounted for only 6.7% of the total sample.

*3.2 Work habits*

About three-quarters (72.8%) of the sample had worked for 10 years or more; just over half (53.1%) were at their desk at least eight hours/day and 79.4% took frequent short breaks. Of those 628 people who took frequent breaks, 14.7% did stretches during their breaks, 10.6% used an app or a timer to set regular reminders and about 4.4% took breaks triggered by pain or discomfort.

*3.3 Exercise outside work*

A total of 43.9% of respondents did less than 2.5 hours of moderate aerobic exercise per week, 33.3% did 2.5-5 hours and 20.1% did more than 5 hours. However, younger respondents took significantly less exercise than their older colleagues (p<0.01) (Figure 1; Table S1).

Chronic conditions greatly limited exercising for 2.5% of respondents.

Less than half the survey population (45%) did muscle-strengthening activities – defined as Pilates, yoga, or lifting weights/children – on at least two days per week.

**Figure 1.** Moderate aerobic exercise in hours per week, percentage by age
Survey question 2.6 *How much moderate\* aerobic exercise do you do on average per week?*

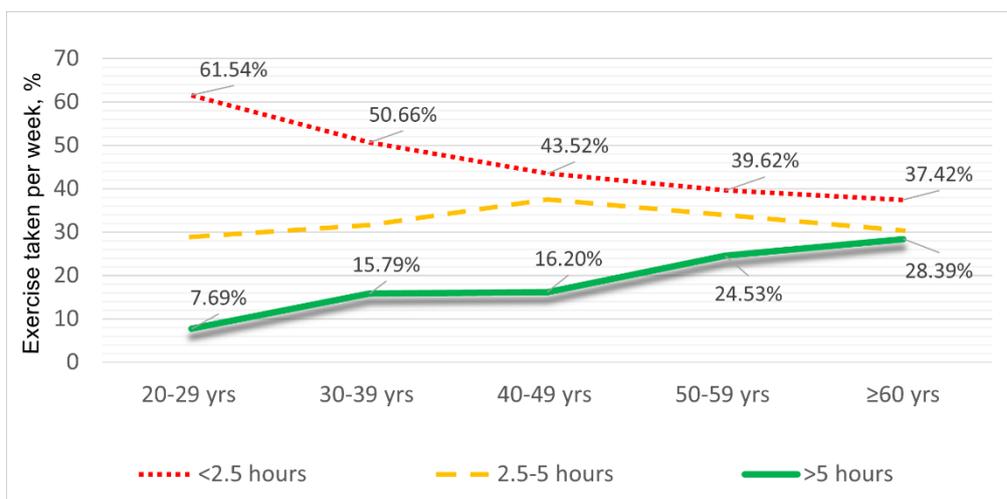

\* Aerobic exercise was defined as *moderate* if the respondent could talk but not sing when doing the activity and *vigorous* if they had difficulty talking when doing the activity. Respondents were instructed to count double the time for vigorous aerobic exercise. Definitions and counting instructions are as per the UK Chief Medical Officers' Physical Activity Guidelines.[5]

*3.4 Upper-body aches and pains*

In the section on pain symptoms – encompassing pain, aches, discomfort, numbness and tingling in the upper body, and excluding headaches or eye problems – point prevalence of pain (upper-body symptoms experienced in the past 7 days) was 43.2% and 12-month prevalence was 69.3%. Overall, 17.2% of respondents had never had any upper-body pain in their working life. However, the mean value per age group revealed disparity, ranging from a mean of only 11.3% in the youngest age group to 25.8% in the oldest age group (Table 1).

**Table 1.** Upper-body pain by age

| Survey question 2.8 *Have you had any aches, pains, discomfort or numbness in your upper body at any time during your working life?* | | | | | | |
|---|---|---|---|---|---|---|
| | 20-29 yrs (n=53) n (%) | 30-39 yrs (n=152) n (%) | 40-49 yrs (n=216) n (%) | 50-59 yrs (n=213) n (%) | ≥60 yrs (n=155) n (%) | Total (n=789) n (%) |
| Yes, including in the past 7 days | 30 (56.60) | 66 (43.42) | 103 (47.69) | 96 (45.07) | 46 (29.68) | 341 (43.22) |
| Yes, in the past 12 months but not in the past 7 days | 14 (26.42) | 46 (30.26) | 62 (28.70) | 53 (24.88) | 31 (20.00) | 206 (26.11) |
| Yes, but not in the past 12 months | 3 (5.66) | 15 (9.87) | 22 (10.19) | 28 (13.15) | 38 (24.52) | 106 (13.43) |
| No | 6 (11.32) | 25 (16.45) | 29 (13.43) | 36 (16.90) | 40 (25.81) | 136 (17.24) |

I found statistically significant differences in 12-month pain prevalence by sex, exercise and age.

Twelve-month pain prevalence was significantly higher among females than males (72.0% vs 55.0%, odds ratio [OR] 0.478, χ² 15.54, degrees of freedom [df] 1, p<0.001) (Table S2).

Respondents who took the most hours of moderate aerobic exercise per week (more than 5 hours) had a 12-month pain prevalence of 56.0%, which was significantly lower than the overall figure of 69.3% (OR 0.569, χ² 16.43, df 2, p=0.004) (Table S3).

As age increased, upper-body pain decreased: 83.0% of respondents in their 20s had pain in the past year versus only 49.7% of the over 60s (OR 0.428, χ² 37.45, df 4, p<0.001) (Figure 2; Table S4).

**Figure 2.** Twelve-month upper-body pain prevalence by age

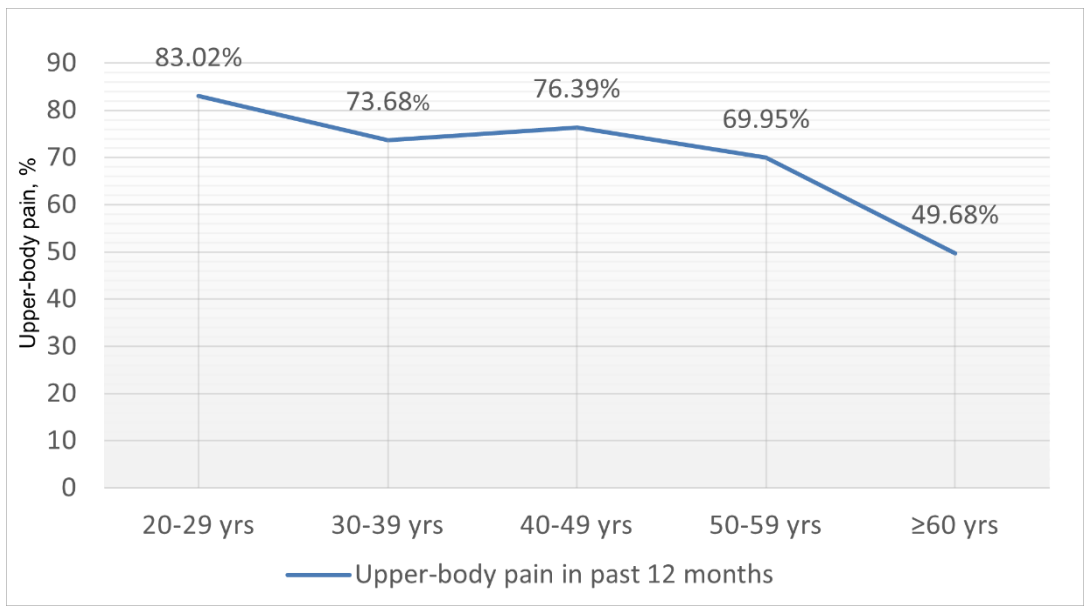

Respondents were asked to select all areas affected by upper-body symptoms during their working life. The most commonly affected body areas were the neck (54.1%), shoulder (52.1%) and lower back (46.0%). Almost half (49.1%) of respondents (n=789) said they had had problems in two or three areas, while about a fifth (21.0%) reported having had problems in four areas or more. These problems were not necessarily synchronous.

When asked which single area had given them the most serious problem, 31.4% of respondents (n=641) reported that the lower back was the most seriously affected area, followed by the shoulder, neck, and wrist/hand. The elbow was cited by far the least (Table S5).

The strategies that most helped resolve lower back pain, when it was cited as the most seriously affected area (n=201), were self-directed targeted exercises and adjustments to posture or existing workstation equipment, mentioned by 47.8% and 36.0%, respectively. Those strategies, closely followed by physiotherapy, were also the top strategies cited to tackle problems in the most seriously affected body areas overall (Figure 3).

**Figure 3.** Survey question 2.11. *Thinking of the most serious problem mentioned above, how did it resolve / how is it improving? Select all measures that you think helped / are helping solve the problem.*

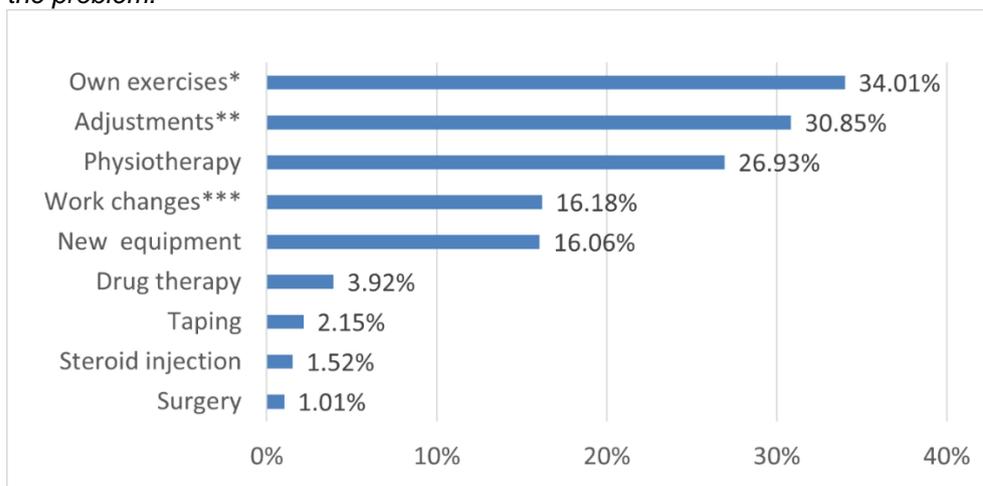

\* self-directed exercises focusing on the most seriously affected area
\*\* adjustments to posture or to existing equipment
\*\*\* changes to working hours or patterns, for which the most common change implemented was taking more frequent breaks

*3.5 Eye problems*

When asked about eye problems (defined as eye strain [tired, itchy or watery eyes], blurred vision, dry eyes and frequent headaches), about a third (33.8%) of respondents reported having symptoms in the past 12 months.

Less exercise and female sex were significantly associated with eye problems ($\chi^2$ 9.37, df 2, p=0.009 and $\chi^2$ 17.32, df 1, p=0.009, respectively). Specifically, 38.7% (OR 0.711) of people who did less than 2.5 hours of exercise and 37.1% of females (OR 0.404) reported eye problems in the past 12 months (Figure 4; Tables S6 and S7).

**Figure 4.** Twelve-month eye-problem prevalence by exercise

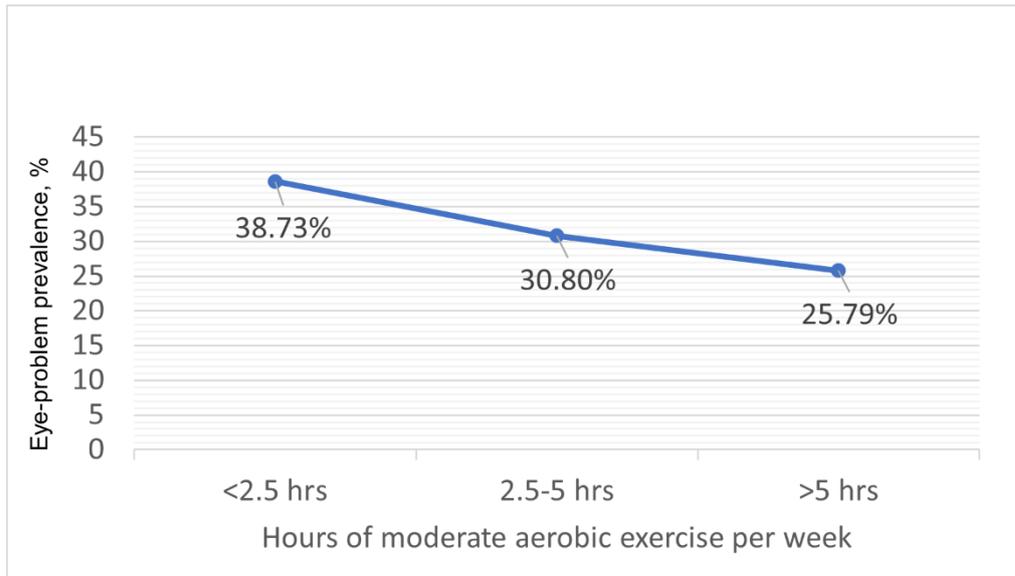

Figure 5 shows that the top strategy that helped alleviate those eye problems was getting new prescription glasses, followed by adjusting the computer monitor (e.g. brightness, contrast) and applying artificial tears.

**Figure 5.** Survey question 4.2.1 *What strategies have you used and found helpful to alleviate your eye problem(s)?*

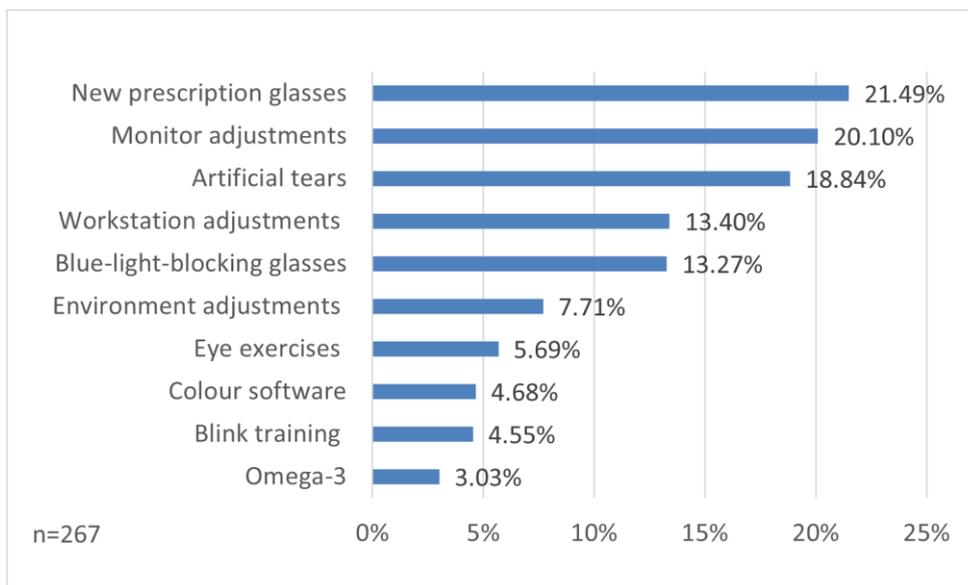

*3.6 Workstation equipment*

I reported on the workstation equipment section of my survey in an [article published in the *ITI Bulletin*](#).[6] To provide readers here with a quick snapshot, the average respondent sits on an IKEA-type office chair with armrests at a fixed-height desk, touch-types on a budget Microsoft keyboard, and uses a Logitech mouse to the right and two monitors in front. The article gives more details on some of the most popular designs, brands and models mentioned in the survey, as well as some anecdotal comments by individual respondents.

# 4. Discussion

*4.1 Demographics*

Language professionals work in a women-dominated industry, as reflected in the 81.3% female respondents to this survey, which was similar to the gender results reported by Cincan in her 2023 Freelance Translator Survey.[7]

Young language professionals were poorly represented in the survey, possibly because many people in their 20s are still studying and those in younger age groups in general may be working in other fields before making a career switch. The high percentage of respondents who had worked for 10 years or more (72.8%), partly impacted by the age disparity, was also reported by the authors of the 2023 ELIS Language Industry Survey,[8] who attributed this seniority to higher affiliation and industry awareness in this demographic. Furthermore, people who are more established in their careers may be more interested in the long-term effects of the sedentary nature of their jobs (while enjoying sufficient financial stability to invest in new ergonomic equipment) and might therefore be more open to completing my survey.

The World Health Organization 2020 Guidelines on Physical Activity and Sedentary Behaviour recommend doing a minimum of 2.5 hours' moderate-intensity exercise per week,[1] and yet 43.9% of survey respondents did less than this, compared with 67% of EU citizens aged 15 and over according to the 2019 Eurostat database.[9] Therefore, language professionals appear to be more active than the general population, despite almost half the respondents not meeting the minimum exercise recommendation.

*4.2 Upper-body pain prevalence*

Pain prevalence in this survey was 69.3%, lower than the 84% figure reported in a New Zealand study in 2009 on MSK disorders among office workers,[10] but similar to the findings of a 2021 study by Holzgreve,[11] who reported pain prevalence figures of 61% (neck), 51% (shoulder) and 48% (lower back), also among office workers.

The higher pain prevalence among females versus males in the survey is widely corroborated by other studies on MSK pain and disorders.[12, 13, 14]

The age association with pain prevalence was surprising, because increasing age appears to be a protective factor in my survey, and yet it is well known that people in the general population experience increasing MSK pain with age.[15, 16, 17] The significantly lower pain prevalence among older respondents may have several explanations. Firstly, these older respondents took more exercise on average than their younger counterparts. Such an association may be due to temporal bias, since older people may exercise more in an attempt to alleviate increasing MSK pain from ageing, i.e. pain onset is prior to exercise and not vice versa. However, vigorous exercise has been associated with a reduced risk of developing MSK pain in longitudinal studies,[18] which makes a temporal bias in my results less likely. A second hypothesis stems from the healthy-worker effect, where people with chronic MSK problems opt for career changes or early retirement, leaving healthier active workers in the older age groups.[19] Thirdly, younger respondents may be making heavier use of electronic devices outside their working hours. This practice was beyond the scope of my survey, but a 2023 systematic review and meta-analysis[20] found that mobile phone use is indeed associated with a considerable risk of neck pain among adults, especially younger adults. This association might explain the high number of respondents in their 20s reporting neck pain versus the entire survey population (67.9% vs 54.1%). Similarly, a study in 2008 on MSK pain prevalence in university employees found higher rates of wrist pain among younger workers than their older counterparts.[14]

Also surprising was the higher proportion of older compared to younger respondents reporting no upper-body pain ever in their working life. I hypothesise that recall bias is likely to have skewed these results, with older respondents not remembering (or not giving importance to) pain experienced decades ago, while respondents in their 20s would have fresh memories of any pain symptoms in their short working life. Intense use of mobile phones mentioned in the paragraph above may also help explain why 88.7% of these young respondents had already experienced upper-body pain.

*4.3 Eye-problem prevalence*

Survey figures were unexpectedly low for eye problems. The 33.8% prevalence in my survey is about half the 66% pooled prevalence reported in a 2022 systematic review and meta-analysis by Lema and Anbesu[2] on computer vision syndrome (CVS), which encompasses all eye and vision problems related to computer use. One explanation for this discrepancy might be the authors' finding that computer users who are well informed about eye problems are at lower risk for CVS. So perhaps language professionals have a lower prevalence of CVS because of their inquisitive nature and well-developed research techniques.

The significant association between female sex and CVS found in this survey is consistent with the fact that females are more at risk for dry eye disease,[21] which is one of the main symptoms of CVS.

People taking more exercise had lower eye-problem prevalence, so they may well be exercising their eye muscles too, changing their focal point from near to middle and long distances. Such focus changes are the basis of the 20-20-20 eye exercise recommended by the American Optometric Association,[22] whereby computer users should take a 20-second break every 20 minutes to view something 20 feet (about 6 metres) away.

Despite the significant association found in this survey between increased exercise and lower prevalence of upper-body and eye problems, readers should note that insufficient exercise does not necessarily cause these problems. Taking more exercise may not protect individuals from experiencing upper-body and eye problems, as many more factors, and their combinations, influence the onset of pain.

# 5. Limitations

Although the survey population was fairly large, it may not accurately reflect language professionals worldwide, particularly those in the younger age bracket who were underrepresented. I adapted the Standardized Nordic questionnaire for MSK symptoms to keep the health section of my survey to a reasonable length, and it was therefore hard to compare my data with other studies that applied the questionnaire in its original format. Similarly, I investigated upper-body pain only, whereas many studies report all-body MSK pain, which makes prevalence rate comparisons less reliable. Despite extensive reading and research, I was unable to find data for MSK pain prevalence in the general population, which would have been particularly useful for comparative purposes.

*Statistical flaws and lessons learned*

The survey had a descriptive design, but collected mainly categorical rather than numerical or continuous data, which made it impossible to perform some statistical tests that might have provided more robust conclusions. For example, because I grouped age by decade, I was unable to identify the exact age range (youngest and oldest respondents) or accurately calculate central tendency; collecting nominal data (e.g. a lot, a little, or no exercise) prevented me from performing the Shapiro-Wilk test to confirm normal distribution; and investigating pain with discrete ordinal responses rather than a Likert scale meant that the Kruskal-Wallis test was less reliable.

Coding categorical data (i.e. assigning numerical values to responses) clearly facilitated the statistical analysis, but, for example, it was unhelpful to code pain as 0 (no pain), 1 (pain but not in past 12 months), 2 (pain in past 12 months but not in past 7 days) and 3 (pain in past 7 days). This was why I grouped pain responses into a dichotomous (yes/no) dataset, as this enabled me to identify 12-month pain prevalence and use chi-squared contingency tables to confirm the statistical significance of associations between independent and dependent variables.

Conducting the pilot survey beforehand was an essential step to ensure adequate content scope and questionnaire readability, but I should have also performed a full data analysis on the pilot responses to detect the above statistical flaws in advance. Hiring a statistician at the pilot stage would have prevented some of these problems, but finding one proved harder than I expected in my capacity as a freelance language professional unattached to a research group or university.

I hope this detailed description of the survey's limitations will be helpful for language professionals and other independent researchers designing surveys in the future.

## 6. Conclusions

Most language professionals appear to be unaffected by eye problems despite working long hours at the computer, possibly because they have effective strategies in place to prevent or alleviate these problems. Upper-body MSK pain is, however, common among language professionals, and female, more sedentary and younger professionals seem to be at greater risk. Further investigation is needed to elucidate the reasons for the discrepancy between the young-age risk factor found in this study and the opposite trend reported in the literature.

## Acknowledgements

I would like to thank Mary Ellen Kerans for her guidance on reporting my survey findings, Marije de Jager and Ailish Maher for editing the article and Hira Mahmood for her help with the statistical analysis. I am also grateful to the participants who gave me helpful feedback on the pilot survey.